\documentstyle[preprint,aps,graphicx]{revtex}
\newcommand{\bold}[1]{\mbox{\boldmath $#1$}}
\newcommand{\dd}{{\rm d}}
\begin{document}
\draft

\title{\flushleft The electromagnetic self-force on a charged spherical body 
slowly undergoing  a small, temporary displacement from a position of rest}

\author{\flushleft V. Hnizdo}

\address{\flushleft National Institute for Occupational Safety
and Health, 1095 Willowdale Road, Morgantown, West Virginia 26505}
\maketitle                                      

\begin{abstract}
The  self-force of classical electrodynamics on a charged ``rigid" body of 
radius $R$ is evaluated analytically for the body 
undergoing a slow (i.e., with a speed $ v\ll c$), slight (i.e., small compared
to $R$), and temporary displacement from an initial position of rest.  
The results are relevant to the Bohr--Rosenfeld analysis of the measurability
of the electromagnetic field, which has been the subject of a recent 
controversy.
\end{abstract}

\begin{flushleft}
\bf 1. Introduction
\end{flushleft}
The problem of the classical electromagnetic self-force on an extended 
charged body moving along a given trajectory is interesting in its own right;
its analysis under some
greatly simplifying conditions is a key ingredient of the well known 
paper on the measurability of the electromagnetic field
of Bohr and Rosenfeld (BR) \cite{BR}. BR derived an 
eight-dimensional-integral expression for
a time average $\bar{F}_{{\rm BR}\,x}$ of the self-force plus the 
electrostatic force due to a stationary
neutralizing body of opposite charge, acting on a charged ``rigid" 
body\footnote{ 
Of course, absolute rigidity is not allowed in a relativistic
theory, and BR went to great lengths to justify an assumption that the body is 
only rigid to the degree that all its parts participate sufficiently uniformly 
in the body's assumed motion.} 
that is undergoing an $x$ direction displacement whose time dependence 
$D_x(t_1)$ approaches sufficiently closely a steplike trajectory 
$D_x\Theta(T-t_1)\Theta(t_1)$, $D_x=\rm const$ [$\Theta(x)$ is 
the Heaviside step function]. The expression of BR is written as 
\begin{equation}
\bar{F}_{{\rm BR}\,x}=\rho_c^2V^2T D_x\bar{A }^{\rm(I,I)}_{xx}
\label{FBR}
\end{equation}
where $V$ and $\rho_c$ are the displaced body's volume and constant
charge density, respectively, and the quantity  $\bar{A}^{\rm(I,I)}_{xx}$ 
is the BR geometric factor
for two fully coinciding space-time regions I of volume $V$ and duration $T$
\cite{BR,VH}:
\begin{equation}
\bar{A}^{\rm(I,I)}_{xx}=-\frac{1}{V^2T^2}\int_{T}\dd t_1\int_{T}\dd t_2
\int_V\dd\bold{r}_1\int_V \dd\bold{r}_2 
\left(\frac{\partial^2}{\partial x_1
\partial x_2}-\frac{\partial^2}{\partial t_1\partial t_2} \right)
\frac{\delta(t-r)}{r}.
\label{AXX}
\end{equation}
Here and henceforth,  $t=t_2-t_1$, $r=|\bold{ r}_2-\bold{ r}_1|$,
and units such that the speed of light $c=1$ are used. 
This result is valid only when the displacement
$|D_x|\ll a$, where $a$ characterizes the linear dimensions of the body,
and the speed\footnote{We shall display 
explicitly the factor $c$ in some inequalities.} 
$|\dot{D}_x(t_1)|\ll c$---which implies a further condition 
that $|D_x|\ll c \Delta t$, where $\Delta t\ll T$ is the duration of the 
time intervals during which the displacement $D_x(t_1)$ goes smoothly 
from zero 
to the constant value $D_x$ and from $D_x$ back to zero at the beginning and 
end, respectively, of the given time period $0\le t_1\le T$.

Recently, Compagno and Persico (CP) \cite{CP} have questioned the 
use of a steplike trajectory in the  
BR calculation of the self-force, and have drawn the conclusion that
the BR result that a single space-time-averaged component of the 
electromagnetic field can be measured to arbitrary accuracy only 
using a compensating spring is incorrect since it is based on
the expression (\ref{FBR}) that assumes an unphysical steplike trajectory. 
The paper of CP is 
criticized in a Comment \cite{Com}, where it is shown by an explicit
calculation that the limiting  BR
time-averaged self-force (\ref{FBR})  approximates correctly 
the self-force obtained with a ``physical" trajectory of 
$|\dot{D}_x(t_1)|\ll c$ 
but with a sufficiently short duration $\Delta t$ of the initial and final 
trajectory segments outside which the body is essentially at rest.
In the Reply of CP \cite{Rep}, this criticism is rejected, claiming that 
the calculation in \cite{Com} is incorrect. 

In the present paper, we obtain analytical expressions for the time dependence 
as well as a time average of the self-force on a spherical charged 
``rigid"\footnote{We employ here the same concept of rigidity as BR (see the
first footnote).} body of radius $R$
moving on a trajectory that is subject to the special BR conditions but is not
necessarily of a steplike character.
Exploiting the spherical symmetry of the problem, we
perform the requisite integrations directly with no recourse to the Fourier 
transform methods used in 
\cite{VH,Com}, but in full agreement with the results obtained 
using the Fourier transform method in \cite{Com} and rejected by CP as 
incorrect. The expressions obtained are relatively simple,
and it is surprising that  
such or similar results do not seem to have appeared in the literature before
(with partial exception of papers \cite{VH,Com})---which perhaps is a factor 
behind the recently expressed reluctance to accept them, and their implications,
as correct.
\begin{flushleft}
\bf 2. The time average of the self-force
\end{flushleft}
First, we outline the derivation of a multidimensional-integral expression for 
the self-force in terms of  
the body's trajectory $D_x(t_1)$ that,
while conforming to the BR conditions $|D_x(t_1)|\ll R$ 
and $|\dot{D}_x(t_1)|\ll c$,
is not necessarily of a steplike  
character---apart from satisfying the condition that $D_x(t_1)=0$ for $t_1<0$ 
and $t_1>T$. A detailed derivation of such an expression  
has been given by CP \cite{CP}---but under the  complicating conditions of
a temporary removal of the neutralizing body, which we shall consider simply
as only permanently absent or present. The displaced body's time-dependent
charge density is approximated to first order in a displacement
$\bold{D}(t_1)$ as
\begin{equation}
\rho(\bold{ r}_1,t_1)=
\rho[\bold{r}_1-\bold{ D}(t_1)]\approx[1-\bold{ D}(t_1)\bold{\cdot\nabla}_1]
\rho(r_1) 
\label{RHO}
\end{equation}
where $\rho(r_1)$ is the body's spherically symmetric charge density
before its displacement. Using this approximation
and placing the differential operator $\bold{ D}(t_1)\bold{\cdot\nabla}_1$
suitably using integration by parts,
the retarded potentials of the electromagnetic self-field of the body
can be expressed to first order in the displacement and neglecting also terms
of order $\dot{D}D$ as
\begin{eqnarray}
\phi(\bold{r}_2,t_2)&=&\int\dd\bold{r}_1\int_{-\infty}^{\infty}
\dd t_1\frac{\rho(\bold{ r}_1,t_1)}{r}\delta(t-r)\nonumber \\
&=&\int \dd\bold{ r}_1  \frac{\rho(r_1)}{r}+\int \dd\bold{ r}_1\rho(r_1)
\int_{-\infty}^{\infty} \dd t_1\bold{ D}(t_1)\bold{ \cdot \nabla}_1
\frac{\delta(t-r)}{r}\label{PHI}\\
\bold{A}(\bold{r}_2,t_2)&=&\int \dd\bold{ r}_1\int_{-\infty}^{\infty} \dd t_1
\frac{\rho(\bold{ r}_1,t_1)\dot{\bold{ D}}(t_1)}{r}\delta(t-r)\nonumber \\
&=&\int \dd\bold{r}_1\rho(r_1)\int_{-\infty}^{\infty} \dd t_1
\dot{\bold{D}}(t_1)\frac{\delta(t-r)}{r}.
\label{A}
\end{eqnarray}
Assuming now that the displacement is along the $x$ direction,
the $x$ component of the body's electric self-field can be written as 
\begin{eqnarray}
E_x(\bold{ r}_2,t_2)&=&-\frac{\partial\phi(\bold{ r}_2,t_2)}{\partial x_2}
-\frac{\partial A_x(\bold{ r}_2,t_2)}{\partial t_2} \nonumber\\
&=&-\int \dd\bold{ r}_1\frac{\partial}{\partial x_2}\frac{\rho(r_1)}{r}
\nonumber \\
&&-\int \dd\bold{ r}_1\rho(r_1)\int_{-\infty}^{\infty} \dd t_1
D_x(t_1)\left(\frac{\partial^2}{\partial x_1\partial x_2}
-\frac{\partial^2}{\partial t_1\partial t_2}\right)\frac{\delta(t-r)}{r}.
\label{EX}
\end{eqnarray}
(The magnetic field is neglected in view of the assumption that 
the body's speed $|\dot{D}_x(t_1)|\ll c$.)
This results in a self-force ${\cal F}_x(t_2)$
on the  displaced body given to first order in $D_x$ by
\begin{equation}
{\cal F}_x(t_2)=\int \dd \bold{ r}_2\left\{\left[1-D_x(t_2)\frac{\partial}
{\partial x_2}\right]\rho(r_2)\right\} E_x(\bold{ r}_2,t_2)
=F_{0\,x}(t_2)+F_x(t_2)
\end{equation}
where 
\begin{equation}
F_{0\,x}(t_2)=-D_x(t_2)\rho_c^2\int_{|\bold{ r}_1|<R} \dd\bold{r}_1
\int_{|\bold{ r}_2|<R} \dd \bold{ r}_2
\frac{\partial^2}{\partial x_2^2}\frac{1}{r}
\label{F0X}
\end{equation}
which equals\footnote{This follows  
 on the replacement of $\partial^2/\partial x^2_2$ in (\ref{F0X}) 
by $\case{1}{3} \nabla^2_2$, 
which is allowed by the spherical symmetry of the problem,
and the use of $\nabla^2_2(1/r)=-4\pi \delta^{(3)}(\bold{r}_2-\bold{r}_1)$.}
 $\rho_c^2V^2D_x(t_2)/R^3$ and is, 
for $|D_x|\ll R$, the electrostatic repulsive force that would be 
due to an identical body placed at the undisplaced position; and
\begin{equation}
F_x(t_2)=
-\rho_c^2\int_{|\bold{ r}_1|<R}\dd\bold{ r}_1\int_{|\bold{ r}_2|<R}
\dd\bold{ r}_2\int_0^{T}\dd t_1D_x(t_1) 
\left(\frac{\partial^2}{\partial x_1\partial x_2}
-\frac{\partial^2}{\partial t_1\partial t_2}\right)\frac{\delta(t-r)}{r}
\label{FX}
\end{equation}
which would be the net force in the presence of an oppositely charged 
neutralizing body
occupying permanently the space region of the undisplaced body\footnote{ 
The bodies may be assumed to have a fine tubular structure
that enables them to move without hindrance through each other along a given
direction (see \cite{BR}).}, as the 
electrostatic force of attraction to the neutralizing body would cancel
the force $F_{0\,x}(t_2)$ of equation (\ref{F0X}).
We assumed in equations (\ref{F0X}) and (\ref{FX}) that the body has a 
constant charge density
$\rho_c$, and replaced the infinite region of the time integration
with the time interval $(0,T)$ in view of the fact that
the displacement $D_x(t_1)\equiv 0$ outside this interval. 
We shall be calling, following CP, the force $F_x(t_2)$ of equation 
(\ref{FX}) also a ``self-force."

We now evaluate the time average of the self-force as a one-dimensional
quadrature involving the body's trajectory $D_x(t_1)$.
A time-averaged self-force $\bar{F}_x$ is obtained by averaging the expression (\ref{FX})
with respect to time $t_2$,
\begin{equation}
\bar{F}_x=\frac{1}{T}\int_0^{T}\dd t_2\,F_x(t_2)
\label{FBARX}
\end{equation}
and we note 
that when the trajectory $D_x(t_1)$ in (\ref{FX}) is replaced formally  by a 
steplike trajectory $D_x\Theta(T-t_1)\Theta(t_1)$, the averaging results
in the limiting BR time-averaged self-force $\bar{F}_{{\rm BR}\,x}$ 
of equation (\ref{FBR}).
The time-averaged self-force (\ref{FBARX}) can be written as 
\begin{equation}
\bar{F}_x=\frac{\rho_c^2V^2}{T}\int_0^{T}\dd t_1\,D_x(t_1)f(t_1)\;\;\;\;\;
V=\frac{4}{3}\pi R^3
\label{FBARXD}
\end{equation} 
where the function $f(t_1)$ is defined by 
\begin{equation}
f(t_1)=-\frac{1}{V^2}\int_{|\bold{ r}_1|<R}\dd\bold{ r}_1\int_{|\bold{ r}_2|<R}
\dd\bold{ r}_2\int_0^{T}\dd t_ 2 
\left(\frac{\partial^2}{\partial x_1\partial x_2}
-\frac{\partial^2}{\partial t_1\partial t_2}\right)\frac{\delta(t-r)}{r}.
\label{FT1}
\end{equation}
We note with CP \cite{Rep} that the function $f(t_1)$ can be written as 
\begin{equation} 
f(t_1)=\frac{1}{V^2}\int_{|\bold{r}_1|<R}\dd\bold{r}_1\int_{|\bold{ r}_2|<R}
\dd\bold{r}_2\int_0^{T}\dd
t_2
\left[\frac{1}{3}\left(\nabla_2^2-
\frac{\partial^2}{\partial t_2^2}\right) +\frac{2}{3}\frac{\partial^2}
{\partial t_1 \partial t_2}\right]\frac{\delta(t-r)}{r}
\label{FT11}
\end{equation} 
because
$\partial/\partial x_1=-\partial/\partial x_2$ and $\partial/\partial t_1=
-\partial /\partial t_2$ when operating on $\delta(t-r)/r$, and
$\partial^2/\partial x_2^2$ can be replaced by $\case{1}{3}\nabla^2_2$ 
in view of 
the spherical symmetry of the problem.
Using the well-known equation for the retarded Green's function 
$\delta(t-r)/r$,
\begin{equation}
\left(\nabla^2_2-\frac{\partial^2}{\partial t_2^2}\right)\frac{\delta(t-r)}{r}
=-4\pi\delta^{(3)}(\bold{ r}_2 -\bold{ r}_1) \delta(t_2-t_1) 
\label{GREEN}
\end{equation} 
and performing also the integration with respect to $t_2$ in the second term 
of (\ref{FT11}), $f(t_1)$ is obtained as 
\begin{eqnarray}
f(t_1)=&&-\frac{1}{R^3}[\Theta(T-t_1)-\Theta(-t_1)]\nonumber \\
&&-\frac{2}{3V^2} \int_{|\bold{ r}_1|<R}\dd\bold{ r}_1\int_{|\bold{ r}_2|<R}
\dd\bold{ r}_2\,\frac{1}{r}[\delta'(T-t_1-r)-\delta'(-t_1-r)].
\label{FT12}
\end{eqnarray} 
Defining an integral 
\begin{equation}
I(s)=\frac{1}{V^2}\int_{|\bold{ r}_1|<R}\dd\bold{ r}_1\int_{|\bold{ r}_2|<R}
\dd\bold{ r}_2 \frac{\delta'(s-r)}{r}\;\;\;\;\;
V=\frac{4}{3}\pi R^3\;\;\;\;r=|\bold{ r}_2-\bold{ r}_1| 
\label{IS}
\end{equation}
the function $f(t_1)$ can now be written as
\begin{equation}
f(t_1)=-\frac{1}{R^3}[\Theta(T-t_1) -\Theta(-t_1)]
-\frac{2}{3}[I(T-t_1)-I(-t_1)].
\label{FT1I}
\end{equation}

Due to the symmetry of the 
problem, the integral (\ref{IS}) reduces to a three-dimensional quadrature: 
\begin{equation}
I(s)=\frac{9}{2R^3}\int_0^{1}\!\dd\zeta_1\,\zeta_1^2 \int_0^{1}\dd\zeta_2\,
\zeta_2^2\int_{-1}^1\!\dd x
\frac{\delta'(\xi-\sqrt{\zeta_1^2+\zeta_2^2-2\zeta_1\zeta_2x})}
{\sqrt{\zeta_1^2+\zeta_2^2-2\zeta_1\zeta_2x}}\;\;\;\:\;\xi=\frac{s}{R}.
\label{IS3}
\end{equation}
Let us do the integration with respect to $x$ first. On the substitution
$\xi-(\zeta_1^2+\zeta_2^2-2\zeta_1\zeta_2x)^{1/2}=y$, this yields
\begin{equation}
\int_{-1}^1 \dd x\,
\frac{\delta'(\xi-\sqrt{\zeta_1^2+\zeta_2^2-2\zeta_1\zeta_2x})}
{\sqrt{\zeta_1^2+\zeta_2^2-2\zeta_1\zeta_2x}}
=\frac{\delta[\sqrt{(\zeta_1-\zeta_2)^2}-\xi]-\delta[\sqrt{(\zeta_1+
\zeta_2)^2}-\xi]}{\zeta_1\zeta_2}.
\label{ISX}
\end{equation}
The integration with respect to a radial variable, say $\zeta_1$,
can be done using the rule
$\delta[f(x)]=\sum_i \delta(x-x_i)/|f'(x_i)|$,
where $x_i$ are the roots of $f(x)=0$; $\delta[f(x)]=0$ when there are
 no real roots of $f(x)=0$.
When $\xi\ge 0$, the roots of
$f_+(\zeta_1)\equiv[(\zeta_1+\zeta_2)^2]^{1/2}{-}\xi$ are
$\zeta_{1\,i}^{(+)}=-\zeta_2\pm \xi$,  and the roots of $f_-(\zeta_1)
\equiv[(\zeta_1-\zeta_2)^2]^{1/2}-\xi$ are $\zeta_{1\,i}^{(-)}
=\zeta_2\pm \xi$, with
$|f'_{\pm}(\zeta_{1\,i}^{(\pm)})|=1$ for all these roots; when $\xi<0$,
 there are no real roots of
$f_{\pm}(\zeta_1)=0$.
We thus obtain
\begin{eqnarray} 
K(\zeta_2,\xi)
&\equiv&\int_0^1 \dd\zeta_1\,\zeta_1^2\frac{\delta[\sqrt{(\zeta_1
-\zeta_2)^2}-\xi]
-\delta[\sqrt{(\zeta_1+\zeta_2)^2}-\xi]}{\zeta_1\zeta_2}\nonumber\\
&=&\Theta(\xi)\int_0^1\dd\zeta_1 \frac{\zeta_1}{\zeta_2}
[\delta(\zeta_1-\zeta_2+\xi) +\delta(\zeta_1-\zeta_2-\xi)
-\delta(\zeta_1+\zeta_2+\xi)-\delta(\zeta_1+\zeta_2-\xi)]\nonumber \\
&=&\frac{\Theta(\xi)}{\zeta_2}[(\zeta_2+\xi)\Theta(1-\zeta_2-\xi)
+(\zeta_2-\xi) \Theta(1+\zeta_2-\xi)]\;\;\;\;\;0<\zeta_2<1.
\label{K}
\end{eqnarray}
Finally, 
\begin{equation}
I(s)=\frac{9}{2R^3}\int_0^1\!\dd\zeta_2\,\zeta_2^2K(\zeta_2,\xi)
=\frac{3}{4R^3}(2-\xi)(2-2\xi-\xi^2)
\Theta(\xi)\Theta(2-\xi)\;\;\;\;\;\xi=\frac{s}{R}
\label{ISC} 
\end{equation}
and using this in equation (\ref{FT1I}), the function $f(t_1)$
is evaluated for $0<t_1<T$ in closed form as
\begin{equation}
f(t_1)=-\frac{1}{R^3}-
\frac{1}{2R^3}(2-\chi)(2-2\chi-\chi^2)\Theta(2-\chi)\;\;\;\;\;\; 
\chi=\frac{T-t_1}{R}\;\;\;\;\;\;0<t_1<T
\label{FT1C}
\end{equation}
which completes the evaluation of the time-averaged self-force (\ref{FBARXD}) as
a one-dimensional quadrature involving the test body's displacement
$D_x(t_1)$.
The BR geometric factor (\ref{AXX}) for coinciding spherical space-time
regions can now be obtained in closed form as
\begin{equation}
\bar{A}_{xx}^{\rm(I,I)}=\frac{1}{T^2}\int_0^{T}\dd
t_1\,f(t_1)
=-\frac{1}{R^4\kappa}-\frac{1}{8R^4\kappa}(4+\kappa)(2-\kappa)^2
\Theta(2-\kappa)\;\;\;\;\;\;\kappa=\frac{T}{R}
\label{AXXC}
\end{equation}
in agreement with equation (100) of \cite{VH}. The terms $-1/R^3$ and
$-1/R^4\kappa$ in expressions (\ref{FT1C}) and (\ref{AXXC}), respectively, 
are due to the
electrostatic force of attraction to the neutralizing body---when the latter
is absent, or one is interested only in the proper self-force itself,
these terms must be subtracted from the above expressions.
According to equation (\ref{AXXC}), the BR geometric factor
 $\bar{A}_{xx}^{\rm(I,I)}$
equals the electrostatic term  $-1/R^3T$ only when the duration of 
the displacement
$T\ge 2R$; the result of CP that $\bar{A}_{xx}^{\rm(I,I)}=-1/R^3T$ for
all values of $T$ (see [5, equation (11)]) was obtained 
by an incorrect use of a Taylor expansion of the derivative of the delta 
function in an integration with finite limits.

We assume that the trajectory 
$D_x(t_1)$ can be expanded about the point $t_1=0$
 as a Taylor series, valid for $0< t_1< T$:
\begin{equation}
D_x(t_1)=\sum_{n=0}^{\infty}\frac{D_x^{(n)}(0^+)}{n!}t_1^n.
\label{DXT}
\end{equation}
This enables us to evaluate analytically the time-averaged self-force
(\ref{FBARXD}) 
in terms of the time derivatives $D^{(n)}_x(0^+)\equiv\lim_{t_1\rightarrow 
0^+}\dd^nD_x(t_1)/\dd t_1^n$
using the closed-form expression (\ref{FT1C}) for the function $f(t_1)$:
\begin{eqnarray}
\bar{F}_x&=&\frac{\rho_c^2V^2}{T}\sum_{n=0}^{\infty}
\frac{D_x^{(n)}(0^+)}{n!}\int_0^{T}\dd
t_1\,t_1^nf(t_1)\nonumber \\
&=&\frac{3\rho_c^2V^2}{T R^2}\sum_{n=0}^{\infty}\frac{R^nD_x^{(n)}(0^+)}
{(n+4)!}\big\{[\kappa^2+2(n+2)\kappa+n(n+3)](\kappa-2)^{n+2}\Theta(\kappa-2)
\nonumber \\
&&-\kappa^{n+4}+(n+3)(n+4)\kappa^{n+2}
-(n+2)(n+3)(n+4)\kappa^{n+1}\big\}\;\;\;\;\kappa=\frac{T}{R}.
\label{FBARXA}
\end{eqnarray}
The electrostatic term $-1/R^3$ in (\ref{FT1C}) contributes here one-third of
the $\kappa^{n+1}$ term, and so 
the time-averaged proper self-force itself (or the time-averaged 
``radiation-reaction" component of the ``self-force" {\cite{CP,Rep}) 
is obtained by replacing $\kappa^{n+1}$ in (\ref{FBARXA}) 
with $\case{2}{3}\kappa^{n+1}$.

Figure \ref{FIG1} exhibits the dependence 
on the displacement duration $T$ of the time-averaged self-force 
$\bar{F}_x$ for a trajectory
$D_x(t_1)=D_x[1{-}\cos(2\pi t_1/T)]\Theta(T{-}t_1)\Theta(t_1)$,
as calculated according to equation (\ref{FBARXA}),
together with that of the limiting time-averaged self-force 
$\bar{F}_{{\rm BR}\,x}$, as given by equations (\ref{FBR}) and (\ref{AXXC}).

\begin{flushleft}
\bf 3. The time dependence of the self-force
\end{flushleft}
Using the results obtained in the course of calculating  the time average
of the self-force,
we can also evaluate analytically the time dependence $F_x(t_2)$ of the 
self-force in terms of the derivatives $D_x^{(n)}(0^+)$ of the body's 
trajectory $D_x(t_1)$. The self-force (\ref{FX})
can be written as
\begin{equation}
F_x(t_2)=\rho_c^2V^2 \int_0^{T}\dd t_1D_x(t_1)g(t_2-t_1)
\label{FXT2D}
\end{equation}  
where the function $g(t)$ is defined by
\begin{equation}
g(t)=-\frac{1}{V^2}
\int_{|\bold{ r}_1|<R}\dd\bold{ r}_1\int_{|\bold{ r}_2|<R}
\dd\bold{ r}_2\left(\frac{\partial^2}{\partial x_1\partial x_2}
-\frac{\partial^2}{\partial t_1\partial t_2}\right)\frac{\delta(t-r)}{r}
\label{GT}
\end{equation}
which differs from the definition (\ref{FT1}) of the function $f(t_1)$ only 
by the absence of the integration with respect to time $t_2$. Thus, using
equation (\ref{FT1I}), 
the function $g(t)$ can be expressed in terms of the derivative of the
integral $I(s)$, and using the expression (\ref{ISC}) for $I(s)$, we get
\begin{equation}
g(t)=-\frac{1}{R^3}\delta(t)-\frac{2}{3}\frac{\dd I(t)}{\dd t}
=-\frac{3}{R^3}\delta(t)+\frac{3}{2R^4}(2-\xi^2)\Theta(\xi)\Theta(2-\xi)
\;\;\;\;\;\xi=\frac{t}{R}.
\label{GTC}
\end{equation}
Only one-third of the $\delta(t)$ term on the right-hand side arises from the 
electrostatic term as the derivative of the step function $\Theta(\xi)$ in
$I(s)$ also contributest.
Using in equation (\ref{FXT2D}) the closed-form expression (\ref{GTC})
for $g(t)$ and the Taylor
expansion (\ref{DXT}) for $D_x(t_1)$ in the non-delta-function term,   
we obtain the following analytical expression for the time dependence 
$F_x(t_2)$ of the self-force: 
\begin{eqnarray}
F_x(t_2)&=&-\frac{3\rho_c^2V^2}{R^3}D_x(t_2)+\frac{3\rho^2V^2}{R^3}\Theta(t_2)
\sum_{n=0}^{\infty}\frac{R^nD^{(n)}_x(0^+)}{(n+3)!}\nonumber\\
&&\times\big[a_n(\kappa)\Theta(\kappa-2)\Theta(2-\xi)
+b_n(\kappa)\kappa^{n+1}\Theta(-\xi)+c_n(\kappa)\Theta(\xi)\Theta(2-\xi)\big]
\nonumber \\
a_n(\kappa)&=&[\kappa^2+2(n+1)\kappa+n^2+n-2](\kappa-2)^{n+1}\;\;\;\;\;
b_n(\kappa)=-\kappa^2+n^2+5n+6\nonumber \\
c_n(\kappa)&=&\frac{T^{n+1}}{R^{n+1}}[b_n(\kappa)-(n+1)\xi\kappa
-\case{1}{2}(n^2+3n+2)\xi^2]\;\;\;\;\;
\kappa=\frac{t_2}{R}\;\;\;\;\;\xi=\kappa-\frac{T}{R}.
\label{FXT2A}
\end{eqnarray}
We note that, interestingly, the electrostatic force
$-\rho_c^2V^2D_x(t_2)/R^3$ of attraction to the neutralizing body 
contributes here only one third of the term that is directly proportional
to the instantaneous distance $|D_x(t_2)|\ll R$ from the neutralizing body.
The averaging of expression (\ref{FXT2A}) according to 
equation (\ref{FBARX}) confirms equation (\ref{FBARXA})
for the time-averaged self-force $\bar{F}_x$; as expected, the self-force 
$F_x(t_2)$  vanishes when the variable $\xi\ge 2$ (i.e., when $t_2\ge T+2R$).
The limiting BR self-force $F_{{\rm BR}\,x}(t_2)$ is obtained with a  steplike trajectory $D_x(t_2)=D_x\Theta(T{-}t_2)\Theta(t_2)$, for which 
only the $n{=}0$ term [with $D_x^{(0)}(0^+)=D_x$] in the series 
in equation (\ref{FXT2A}) is nonzero.

Figures \ref{FIG2} and \ref{FIG3} show the time dependence of the self-force 
$F_x(t_2)$, 
calculated using equation (\ref{FXT2A}) for the trajectory
$D_x(t_2)=D_x[1{-}\cos(2\pi t_2/T)]\Theta(T{-}t_2)\Theta(t_2)$
and for the limiting steplike trajectory $D_x\Theta(T{-}t_2)\Theta(t_2)$.

An alternative expression for the self-force $F_x(t_2)$
in terms of the derivatives $D^{(n)}_x(t_2)$ at a current time $t_2$ along
the trajectory should be instructive since the radiation-reaction  force
is usually expressed in terms of such derivatives. 
This can be done easily by suitably changing the integration variable 
in equation (\ref{FXT2D}) before expanding the trajectory in a Taylor series:  
\begin{equation}
F_x(t_2)
=\rho_c^2V^2\sum_{n=0}^{\infty}\frac{(-1)^n}{n!}D_x^{(n)}(t_2)
\int_{t_2-T}^{t_2}\dd t\,t^ng(t).
\label{FXT2}
\end{equation}
However,  this is valid only for $t_2< T$, as the function 
$D_x(t_2)\equiv 0$ for
$t_2> T$ and as such cannot be expanded about a point $t_2>T$ for use in
the interval $(0,T)$.
The integration in 
equation (\ref{FXT2}) with the closed-form expression (\ref{GTC}) for the
function $g(t)$ leads to the following result:  
\begin{eqnarray}
F_x(t_2)&=&-\frac{3\rho_c^2V^2}{R^3}D_x(t_2)
-\frac{3\rho_c^2V^2}{2R^3}
\Theta(t_2) \sum_{n=0}^{\infty}\frac{(-1)^nR^nD_x^{(n)}(t_2)}{(n+3)(n+1)!}
\nonumber \\
&&\times\{[2^{n+2}(n-1)-d_n(\kappa)] \Theta(\kappa-2)+d_n(\kappa)\}\nonumber\\
d_n(\kappa)&=&[(n+1)\kappa^2-2n-6]\kappa^{n+1}\;\;\;\;\;
\kappa=\frac{t_2}{R}<\frac{T}{R}.
\label{FXT2AA}
\end{eqnarray}
For times $t_2>2R$ (i.e., for $\kappa>2$), equation (\ref{FXT2AA}) gives 
\begin{equation}
F_x(t_2)
=-\frac{\rho_c^2V^2}{R^3}D_x(t_2)-\frac{24\rho_c^2V^2}{R^3}
\sum_{n=0}^{\infty}\frac{(-2)^nR^{n+2}D^{(n+2)}_x(t_2)}{(n+5)(n+3)(n+2)n!}
\;\;\;\;\;2R<t_2<T.
\label{FXT2RR}
\end{equation}
Here, the original $n{=}0$ term reduced the first term of equation 
(\ref{FXT2AA}) to the electrostatic force of attraction to the neutralizing
body, the original $n{=}1$ term vanished, and the summation of the series
was relabeled so that it now begins with the $D^{(2)}_x(t_2)$ term.
The series in equation (\ref{FXT2RR}) agrees with the expression  
given by Jackson \cite{Jack} for the electromagnetic self-force\footnote{
Note that Jackson's self-force is
defined so that its sign is opposite to ours.} 
on a body carrying a spherically
symmetric charge distribution. This can be seen on noting that, 
in the case of a uniform spherically symmetric charge
density, the integral appearing in that expression has the following 
value:
\begin{equation}
\int_{|\bold{ r}_1|<R}\!\dd \bold{ r}_1\int_{|\bold{ r}_2|<R}\!
\dd\bold{ r}_2\,r^{n-1}=\frac{9V^22^{n+2}R^{n-1}}{(n+5)(n+3)(n+2)}.
\label{RNI1}
\end{equation}
This integral was evaluated by reducing it to a three-dimensional quadrature
in the same way as that of the reduction of integral (\ref{IS}) to integral
(\ref{IS3}) and performing the resulting three-dimensional integral 
analytically.

In conclusion, we remark that the fact that the time-averaged self-force
(\ref{FBARXD})
is proportional to the displacement $D_x$ even in the absence of the
neutralizing body---for a displacement duration $T<2R$ and in the
limit of a steplike trajectory---does not contradict
the translational invariance of the Lagrangian of the system consisting of
the displaced body and the electromagnetic field. Such  invariance 
is irrelevant to the case under the consideration
because the body is assumed to be displaced by
an external force whose origin is outside this system.


\begin{figure}
\centering
\includegraphics[angle=270,width=\linewidth, clip]{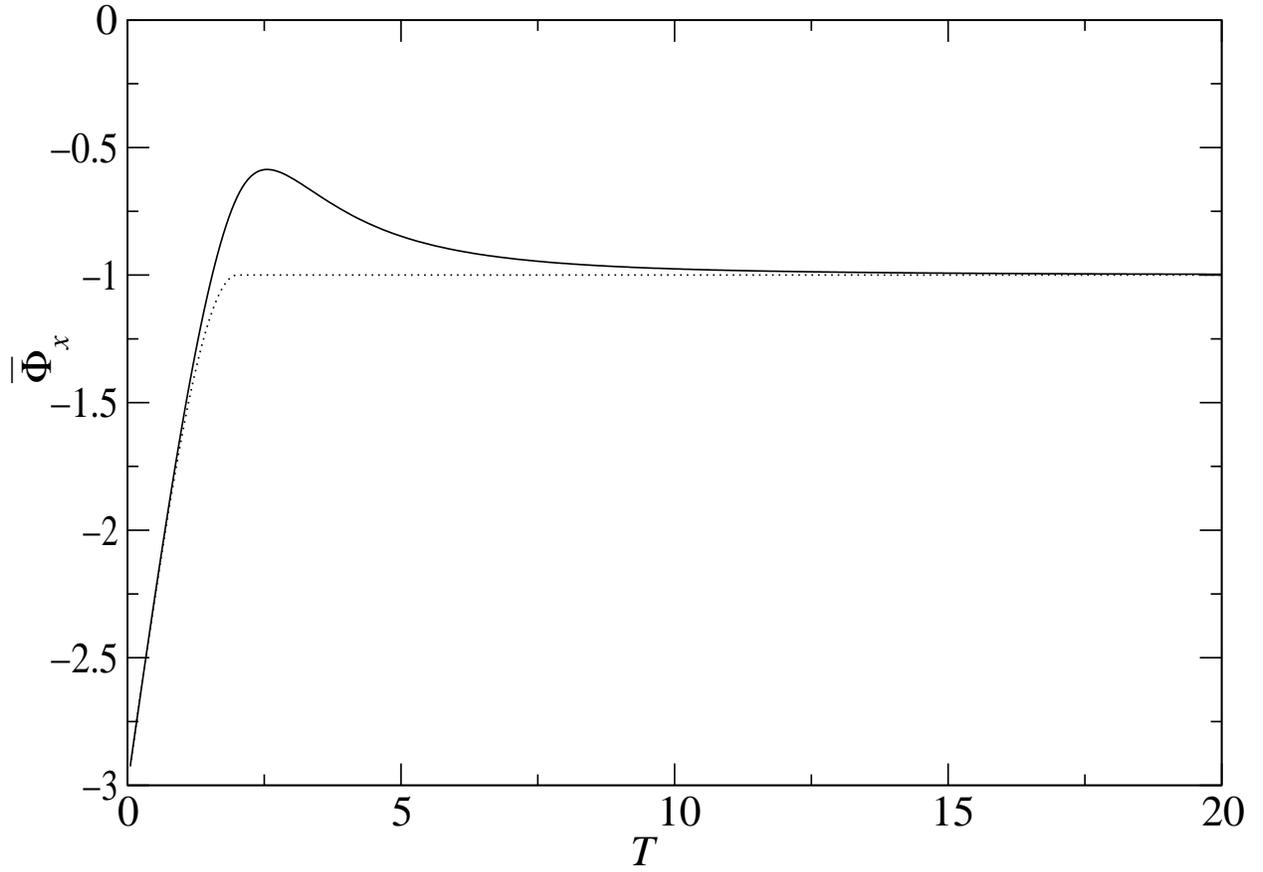}
\vspace{1ex}
\caption
{The normalized time-averaged self-force $\bar{\Phi}_x$ $\equiv$ $\bar{F}_x/
\rho_c^2V^2D_x$ calculated using equation (\ref{FBARXA}) for a trajectory
$D_x(t_1)$ = $D_x[1-\cos(2\pi t_1/T)]\Theta(T-t_1)\Theta(t_1)$
(solid curve), and using equations (\ref{FBR}) and (\ref{AXXC}) for the 
steplike trajectory
$D_x(t_1)$ = $D_x\Theta(T-t_1)\Theta(t_1)$ (dotted curve).
Units such that the speed of light $c=1$ and the radius of the body $R=1$
are used.}
\label{FIG1}
\end{figure}
\begin{figure}
\centering
\includegraphics[angle=270,width=\linewidth, clip]{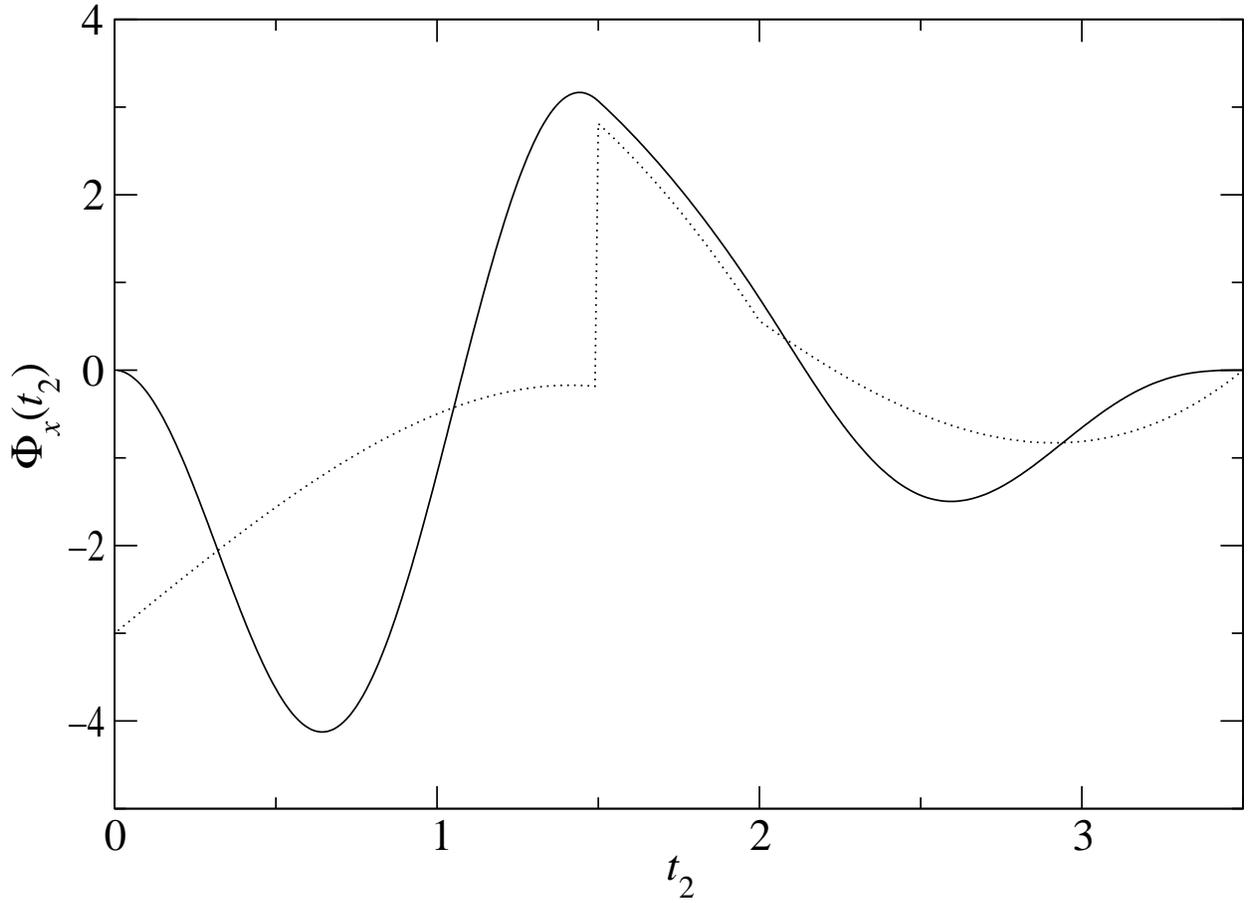}
\vspace{1ex}
\caption
{The normalized  self-force $\Phi_x(t_2)\equiv F_x(t_2)/
\rho_c^2V^2D_x$ calculated using equation (\ref{FXT2A}) for a trajectory
$D_x(t_2)$ = $D_x[1-\cos(2\pi t_2/T)]\Theta(T-t_2)\Theta(t_2)$
(solid curve) and for the steplike trajectory
$D_x(t_2)$ = $D_x\Theta(T-t_2)\Theta(t_2)$ (dotted curve), with $T=1.5$.
Units as in figure \ref{FIG1} are used.}
\label{FIG2}
\end{figure}
\begin{figure}
\centering
\includegraphics[angle=270,width=\linewidth, clip]{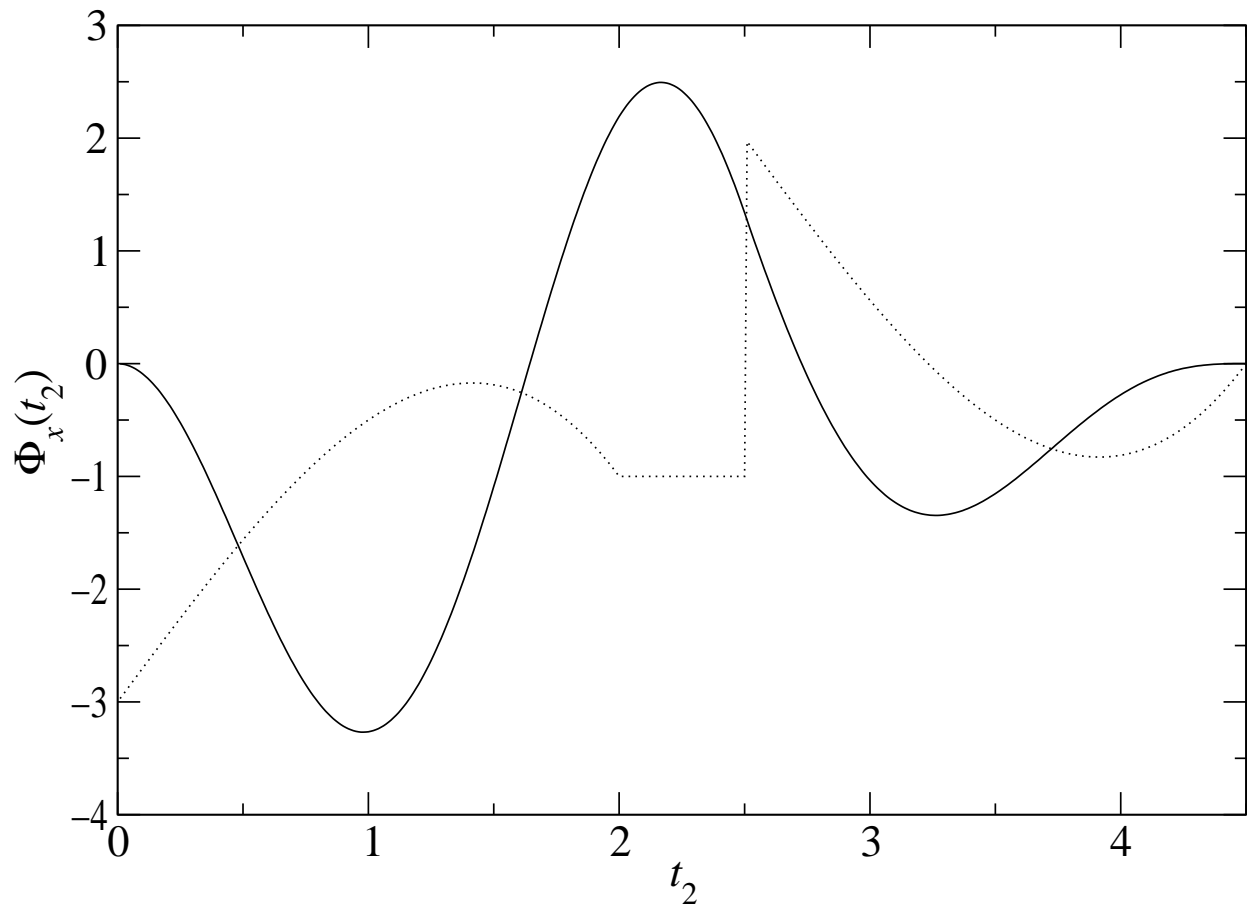}
\vspace{1ex}
\caption
{The normalized  self-force $\Phi_x(t_2)$ as in figure \ref{FIG2} but for a 
displacement duration $T=2.5$.}
\label{FIG3}
\end{figure}

\end{document}